\documentstyle[11pt]{article}
\def \rp { \ \rho_{\varphi} \ }
\def \pp { \ p_{\varphi} \ }
\def \rr { \ \rho_{r} \ }
\def \pd { \ \dot{\varphi} \ }
\def \pdd { \ \ddot{\varphi} \ }
\def \pds { \ {\dot{\varphi}}^2 \ }
\def \pdp { \ {{\varphi}'}^2 \ }
\def \be  {\begin{equation}}
\def \ee  {\end{equation}}
\def \str {\rightarrow}
\def \srl {\sqrt{\lambda}}
\begin{document}
\begin{titlepage}
\[ \]

\begin{center}
{\Large \bf  Inflationary cosmology with scalar field and
                    radiation}
\[ \]
\[ \]
Alexei V. Nesteruk\footnote{School of Computer Science and
Mathematics, Portsmouth University, PO1 2EG England}
\end{center}
\[ \]
\[ \]
{\em Running head:} Inflationary cosmology with scalar field and
                    radiation
\[ \]
\[ \]
\[ \]
{\em Correspondence:}
\[ \]
{\sf A V Nesteruk\\
School of Computer Science and Mathematics\\
Portsmouth University\\
Mercantile House\\
Hampshire Terrace\\
Portsmouth  PO1 2EG\\
England
\[ \]
TEL:   + (01705) 843 108\\
FAX:   + (01705) 843 106\\
EMAIL: nesteruk@sms.port.ac.uk}
\end{titlepage}
\[ \]
\[ \]
{\bf Abstract}
\[ \]
\[ \]
We present a simple, exact and self-consistent cosmology with
a phenomenological model of quantum
creation of radiation due to  decay of the scalar field.
The decay drives a non-isentropic
inflationary epoch, which exits smoothly
to the radiation era, without reheating.
The initial vacuum for radiation is a
regular Minkowski vacuum.
The created radiation obeys standard
thermodynamic laws, and the total entropy produced
is consistent with the accepted value. We analyze the difference
between the present  model and the model with the decaying cosmological
constant considered in \cite{gmn}.
\[ \]
\newpage

\section{Introduction}

The aim of this paper is to extend the scenario of the evolution of
the universe with smooth exit from inflation, and particle production at
the expense of the decaying cosmological constant, developed in
paper \cite{gmn}. In that paper thermodynamics and Einstein's
equations led to an equation in which the Hubble rate $H$ is determined
by the particle number $N$. The model is completed by specifying
the particle creation rate $\Gamma=\dot{N}/N$, which led to a second-order
evolution equation for $H$.
The evolution equation for $H$ then has a remarkably simple exact
solution, in which a non-adiabatic inflationary era exits smoothly
to the radiation era, without a reheating transition.
For this solution, there were given  exact expressions
for the cosmic scale factor, energy density of radiation and vacuum,
temperature, entropy and super-horizon scalar perturbations.

In this paper we would like to generalize the abovementioned
results for the case of a scalar field $\varphi$ interacting with
radiation via the gravitational field, leading to
cosmological evolution with  smooth exit from inflation. Our
particular task is to determine whether the theory formulated in
terms of the scalar field $\varphi$ leads to any new results in
comparison with the previous case of a decaying cosmological
constant.

The background of this paper is constituted by two ideas in
physical cosmology. One of them is connected with the longstanding
attempt to explain all matter in the universe as produced by
quantum creation from vacuum. This has been studied via quantum
field theory in curved spacetime
(see for example \cite{hupark77,creation}).
Most cosmological models exhibit a singularity which
presents  difficulties  for interpreting quantum effects, because
all  macroscopic parameters of created particles are infinite
there.  This leads to the  problem of the initial vacuum
(see \cite{gmn}). One attempt to
overcome these problems is via
incorporating the effect of particle creation
into Einstein's field equations. For example, in
the papers of the Brussels group \cite{edgard1},
the quantum effect of particle creation  is considered
in the context  of the thermodynamics of open systems,
where it is interpreted as an additional negative
pressure, which emerges from a re-interpretation of
the energy-momentum tensor.
This effect is irreversible in the sense that spacetime
can produce matter, leading to growth of entropy,
while the reverse process is thermodynamically
forbidden.

The main difference with our present paper is that
in \cite{edgard1} the law of massive particle creation,
i.e. the mechanism
of energy flow from the gravitational field to matter,
leads to a non-zero number of
particles at the beginning of expansion, described as a fluctuation
of the regular vacuum.
These results were
recently generalized in  a covariant form in \cite{lima1}.
Our approach differs from that of \cite{edgard1,lima1}
in that we do not modify the field equations.
Instead, we associate  the source of created particles
as a decaying vacuum of the inflaton field $\varphi$.

A number of decaying vacuum models has appeared in the literature
(see \cite{lima2,vl,mubarozer,ozertaha} and references cited
there). A review of the different phenomenological
models  of  evolution with variable cosmological term
can be  found in  paper \cite{overcoop}.

Inflationary models with fixed cosmological constant and cold dark
matter have been successful in accounting for the microwave
background and large-scale structure observations, while also
solving the age problem (see \cite{krausturner}). However, these
models are challenged by the reduced upper limits on $\Lambda$
arising from the Supernova Cosmology Project, and also by the
long-standing problem of reconciling the very large
early-universe vacuum energy density with the very low
late-universe limits \cite{vl}.

One resolution of these problems  is a decaying cosmological
constant $\Lambda$ which is treated as a  dynamical parameter.
This approach was typical for the quantum field theorists for
many years (see for example \cite{qft}).
Anything which contributes to the energy density
$\rho_v$ of the vacuum behaves like a cosmological term
via $\Lambda_v=8\pi G \rho_v$. Many potential sources of fluctuating
vacuum energy have been identified (including scalar fields
\cite{Scalar}) which were to give rise to a negative energy
density which grows with time, tending to cancel out any pre-existing
positive cosmological term and drive the net value of $\Lambda$ toward
zero. Processes of this kind are among the most promising ways to
resolve the longstanding cosmological `constant' problem
(see \cite{wei89} for review).
It is worth mentioning the recent paper of Parker \cite{parker}
indicating an attempt to revive  the idea of
the cosmological constant as a purely quantum effect associated with
the renormalization of the general relativistic action.

In ad hoc
prescriptions, the functional form of $\Lambda(t)$ or
$\Lambda(a)$ or $\Lambda(H)$ (where $a$ is the scale factor and
$H$ is the hubble rate) is effectively assumed a-priori
(see the review \cite{overcoop} where all known forms
for $\Lambda$ are listed).
Typically, the solutions arising from ad hoc prescriptions for
$\Lambda$ are rather complicated, and moreover, it is often
difficult to provide a consistent simple interpretation of the
features of particle creation, entropy and thermodynamics.

In common with \cite{lima2} and extending
\cite{gmn}, we attempt to provide some clear and
consistent physical motivation for the particular form of vacuum
decay for the field $\varphi$.
In contrast to many other models, we propose a simple, exact and
thermodynamically consistent cosmological history. The latter
originates from a regular initial vacuum.
Together with  naturally defined asymptotic conditions for the
number of created particles and the form of the potential
for the field $\varphi$ this leads to a simple expansion law and
thermodynamic properties, and to a definite estimate for the
total entropy in the universe.

Non-adiabatic inflationary models
differ from the standard models (see for example
\cite{kolb&turner90}), in that: (a) radiation is created continuously
during inflation, rather than during reheating; (b) the
continuous vacuum decay itself
initiates a smooth exit from inflation to the radiation era; (c)
entropy and heat production take place continuously, without
the need for reheating. In the standard approach,
the scalar field drives
adiabatic (i.e., isentropic)
inflation, followed by a non-equilibrium reheating era
when the field decays into radiation and inflation is brought to an
end. The potential of the field is then the key ingredient.
We argue in this paper that  the potential
(treated usually  as a self-interacting potential) actually
consists of two parts. One of them is the interaction of the field with
radiation via the gravitational field. This term is expressed in purely
geometrical terms and in the case when the field
is zero  plays the role of the decaying cosmological
constant as in  paper \cite{gmn}.  The second part of the
potential, corresponding to self-interaction of the field, is
postulated in a simple form consistent with the standard requirements
of inflationary cosmology.  One should mention that our ansatz of
decomposing  the potential into two parts is in some sense similar to
the postulate of the interaction term in the form $\Gamma\pds$
which was used e.g. in the papers
\cite{mubarozer,yoko,oliveira}. In these papers the strategy was to
modify the second order Klein-Gordon equation.
In our case the decomposition of the potential allows one
to formulate a differential equation of the first order
for the field $\varphi$.

Since the exit from inflation to the radiation era is smooth,
we avoid the problem of matching at the transition.
A similar
smooth evolution has been used in
\cite{roydavid,alexei_96,caldwell}, but
in the context of adiabatic
inflation, and without a consistent physical foundation.
In analogy with \cite{gmn} we show that the ad hoc form of $H(a)$
given in \cite{roydavid} follows from simple physical conditions.
In \cite{b1}, a
kinematic analysis is given for various
non-adiabatic inflationary evolutions with smooth exit, but
these evolutions are outside the scope of our model.

The choice of $a$ as dynamical variable and the very simple form of
$H(a)$ that meets the physical conditions, lead
to elegant expressions for all parameters
describing the radiation and decaying vacuum, and also to a
physically transparent interpretation of these results, including
the estimate of entropy.

We use units with $8\pi G$, $c$ and $k_{\rm B}$ equal 1.

\section{Model}  %
Consider a spatially flat FRW universe
\begin{equation}
ds^2 = - dt^2 + a^2(t) \left[dx^2 + dy^2 + dz^2\right]\,,
\label{metric}
\end{equation}
containing radiation with energy
density $\rho_{r}(t)$ and pressure $p_r = \rho_r/3$.

The energy momentum tensor of these components correspondingly is
\[
T^R_{\mu\nu}={1\over3}\rho_r(t)\left[4u_\mu u_\nu+g_{\mu\nu}\right]\,
\]
We also consider quantum matter described by the scalar field $\varphi$
with energy density and pressure given in a conventional form
\[
- T^{Q}_{0\,0} = \rp = \frac{1}{2} \pds + V, \qquad
T^{Q}_{(i)\,(i)} = \pp =  \frac{1}{2} \pds - V
\]
where $V$ is a potential. In a special case of quantum matter
with the equation of state $\pp = - \rp$, we have $\pds = 0$, and
the potential can be treated as a cosmological constant, dependent on $t$.
\section{Conservation Law}
The conservation equations
$\nabla^\nu(T^R_{\mu\nu} + T^Q_{\mu\nu})=0$
in the general case for $\pds \not = 0$
reduce to
\begin{equation}
\dot{\rr}   +  4 H \rho_r  =  -  \dot{\rp} -  3 H \pds
\label{conserv}
\end{equation}
where $H=\dot{a}/{a}$ is the Hubble rate.
The equation (\ref{conserv})
shows how energy is transferred from the quantum field to radiation.
This energy transfer can be understood as  creation of
quanta of radiation at the expense of the decaying
field $\varphi$. Employing
the extended form of the first law of thermodynamics suggested in
\cite{edgard1} and applied to radiation only:
\begin{equation}
d(\rr {\cal V}) + p_r d{\cal V} -  \frac{h_r}{n_r} d(n_r {\cal V}) = 0 \,,
\label{firstlaw}
\end{equation}
where $h_r = \rr + p_r = 4/3 \rr $,
one can link the evolution of $\rr$ with the
change of the total number of photons $N_r = n_r {\cal V}$, where
${\cal V}$ is the comoving volume of the observable universe.
Since (\ref{firstlaw}) is equivalent to
\begin{equation}
\dot{\rr} + 4 H \rr =  \frac{4}{3} \rr \frac{\dot{N_r}}{N_r} ,
\label{firstlaw1}
\end{equation}
a comparison with (\ref{conserv}) gives for the particle creation rate
\begin{equation}
\Gamma \equiv \frac{\dot{N_r}}{N_r} = -  \frac{\dot{\rp} + 3 H \pds}{4/3 \rr} \,.
\label{evolN}
\end{equation}
The creation of radiation implies that $\dot{N_r}/{N_r} > 0$,
which means in turn that subject to $\rr > 0$, one must have
$\dot{\rp} + 3 H \pds \leq 0$,
i.e.  the energy of the scalar field $\varphi$ decreases in
time, being converted thereby into  quanta of  radiation.

In the case $\pds = 0$, $\rp = V \equiv \Lambda$ so that the equation
(\ref{evolN}) becomes
\begin{equation}
\frac{\dot{N_r}}{N_r} = -  \frac{3\dot{\Lambda}}{4 \rr} \,.
\label{evolNl}
\end{equation}
(this case is considered in \cite{gmn}) i.e. the  production of
classical photons  is a result of decay of the vacuum with the
energy density  $\Lambda$.

If $\pds =0$ and $V=const$ (or $ \Lambda = const$)
the initial vacuum for photons (where $N_r = 0$) is stable leading to no
particle  production. The total number of particles is conserved  and equals
zero.

By switching on the source of radiation $\dot{\rp} \not = 0$ in the
right-hand side of the equations (\ref{conserv}) and
(\ref{evolN}) we effectively switch on the
coupling of radiation with gravitational field,
leading to creation of photons  from vacuum.
\section{Field equations}
The field equations
\[
R_{\mu\nu}-{1\over2}Rg_{\mu\nu}
=T^R_{\mu\nu} + T^Q_{\mu\nu}
\]
are
\begin{eqnarray}
3 H^2 &=& \rho_r  + \rp \,, \label{einst0}\\
2 \dot{H} + 3 H^2 &=& - {1\over3}\rho_r  - \pds + V \,,
\label{einst1}
\end{eqnarray}
and if both are satisfied then the energy equation (\ref{conserv})
follows identically.

Combining (\ref{conserv}) and (\ref{einst0}) one finds
\begin{eqnarray}
\label{rad}
\rr = - \frac{3}{4} [ 2 \dot{H} + \pds ],\\
\label{rp}
\rp = \frac{3}{4} [ 2 \dot{H} + \pds ] + 3 H^2 = \frac{3}{4} \pds
+ \frac{R}{4},
\end{eqnarray}
where $R = 6 \dot{H} + 12 H^2$ is the Ricci curvature.
It is easy to see that the equation (\ref{rp})
is equivalent to the equation linking the potential $V$
and $\pds$:
\be
V =   \frac{1}{4} \pds
+ 3 H^2 + \frac{3}{2} \dot{H}  =  \frac{1}{4} \pds + \frac{R}{4}
\label{potential}
\ee
We can use the equation (\ref{evolN}) in order to express $N_r$
in terms of $\rr$, and then in terms of the combination
 $2\dot{H} + \pds$:
\be
\rr = C \frac{N^{4/3}}{a^4}, \qquad
N = C \left[-\frac{3}{4} (2 \dot{H} + \pds)a^4\right]^{3/4}.
\label{num}
\ee

The formulas (\ref{rad}, \ref{rp}, \ref{num}) give the expressions for
the physical quantities describing matter in the universe in terms of
$\dot{H}$ and $\pds$, which have to be determined from two additional
conditions. First we need the equation
for $\varphi$ and then we need another constraint in order to define
$\dot{H}$.

The standard procedure  to find $\varphi$, and hence $\pds$, is to
obtain  the analogue of the Klein - Gordon equation for the
field $\varphi$.  Indeed, substituting
$\rr$   (\ref{rad}) and $\rp$ (\ref{rp}),
into the equation (\ref{evolN})
one can get the equation
\be
\pdd \pd + \dot{V} + 3 H \pds = [2 \dot{H} + \pds]
\Gamma.
\label{klgor1}
\ee
The difference between this equation  and the conventional form of the
Klein-Gordon  equation (see for example \cite{mubarozer,oliveira})
is that we can not cancel $\pd$ and
proceed to an equation of second order for $\varphi$ because of
the geometrical term $2 \dot{H}\Gamma$ at the right hand side of this
equation. Using the expression for $V$  \  (\ref{potential}) the last
equation can be rearranged in the following form
\be
\frac{3}{2} \pdd \pd + [3 H - \Gamma ]  \pds =
2 \dot{H} \Gamma  -  \frac{\dot{R}}{4}
\label{klgor2}
\ee
From here one can easily realize that the standard
Klein-Gordon  equation  for $\varphi$ will be achievable
only in the case when the right hand side of (\ref{klgor2})
equals zero.

It is interesting to see that in the case  $\pds =0$ the right
hand side of the last equation will be zero anyway, giving the
connection formula between the geometrical parameters of the model
and the rate of particle creation:
\be
2 \dot{H} \Gamma(a)  -  \frac{\dot{R}}{4} =0
\label{connHG}
\ee
This equation has already been obtained in our previous model
\cite{gmn} (equation (9)). It was solved there with respect to
$H$ as  a function of $a$ under  some physical assumptions
about  the rate of particle creation $\Gamma$. This led to a scenario
with a smooth exit from inflation to radiation dominated era
(see details in \cite{gmn}). In the case of the paper \cite{gmn}
it was sufficient to  impose one physical condition
for $\Gamma$ in order to determine $H(a)$.

It is interesting however in the present situation that one can
get a nontrivial solution for $\pd \not = 0$  if the condition
(\ref{connHG}) holds. In this case we have
\be
\frac{d\pds}{dt} +  \left[4 H - \frac{\dot{R}}{6\dot{H}}\right]\, \pds = 0.
\label{klgor3}
\ee

Following \cite{roydavid,gmn}, we use $a$ as a dynamic
variable instead of $t$, and consider  the Hubble rate as
$H=H(a)$ (in this case we can not consider $a=$ constant as a
limiting case for a flat universe). Rewriting the last equation
in terms of $a$, and using the prime as a notation for the
derivative with respect to $a$ ($d/dt = a H d/da = a H ( )'$)
one can rewrite the equation (\ref{klgor3}) in the form
\be
\frac{d\pds}{\pds} =
\left[ \frac{1}{a} + \frac{H'}{H} +
\frac{H''}{H'} \right] \, da
\label{klgor5}
\ee
This equation integrates
to
\[
 \pds = A (- a H H') = A (-\dot{H})
\]
Substituting this result into the formula for $\rr$ one gets
\[
\rr = \left( \frac{3}{2} - A^2 \right) \ (- \dot{H})
\]
which implies that  $A^2 < 3/2$. This formula
together with the condition (\ref{connHG}) reproduces the result
of the paper  \cite{gmn}.

Coming back to the general case  of the equation (\ref{klgor2}),
\be
\dot{Z} + P(H,\Gamma) Z =  Q(H,\Gamma)
\label{klgor6}
\ee
where $Z \equiv \pds$, $P = 4 \ H
-  4/3 \Gamma, Q = 4/3 (2 \dot{H} \Gamma - \dot{R}/4)$,
one can  simplify it using the following substitution
\[
Z = Z_{h} - 2 \dot{H},
\]
which gives the equation for $Z_{h}$ (compare with the equation
(\ref{klgor6}), when $Q=0$):
\[
\dot{Z}_{h} + P(H,\Gamma) Z_{h} = 0
\]
This equation can be easily solved in terms of $N$:
\[
Z_h(a) = C N^{4/3}(a) a^{-4}, \; {\rm so \; \, that} \;
Z \equiv \pds = - 2 \dot{H} +  C N^{4/3}(a) a^{-4}
\]
(where $C$ is an arbitrary constant) which is entirely
equivalent to (\ref{num}). In other words in order to determine
$\pds$ one needs to know $N$ and $H$, and this shows that
we have exhausted all information about our system from the
conservation laws (\ref{conserv}), (\ref{evolN}) and the field
equation (\ref{einst0}). One should develop some further arguments
in order to determine $\pds$ and $H$.
\section{The potential}
The expression for the potential $V$ which we obtained above
(\ref{potential}) is very interesting because it consists of two
parts, one  proportional to the  kinetic energy of the field
$\pds$,  another  explicitly a  function of time or $a$,
which describes the change  of energy in the system field-
radiation due to  expansion of the universe.
This makes it possible to argue that the potential can be
interpreted in the following way:
\be
V(\varphi, a) = U(\varphi) +  V_{int}(a),
\label{potential2}
\ee
where
\be
U(\varphi) = \frac{1}{4} \pds, \qquad  V_{int}(a) =  \frac{R(a)}{4}
\label{potu}
\ee

The potential $U$  describes the self-interaction potential
of the field $\varphi$, whereas $V_{int}$ corresponds to interaction
of the field $\varphi$ with radiation  via the gravitational
field.  In the case of no gravitational field, i.e. $H=0$,
$V = 0$.

Now we are in a position to solve equation  (\ref{potu})
with the initial condition $U \str 0$ as $a \str 0$  (which
follows from (\ref{rad}) as an initial condition for
$\pds (a=0) =0$). The solution  depends, obviously, on the
form of $U$. The contribution of $\pds$ in the formula for
$\rr$ (\ref{rad}) can be treated in this case as an additional
external  source of radiation. Solving the equation (\ref{potu})
does not give, however, any information about $H(a)$, so that we still
need some additional physical arguments in order to fix $H(a)$.

The choice of $U$ is not a trivial one  in our case. Since we would like
to develop a scenario of the evolution of the universe with  inflation it is
natural to assume that  the potential can be chosen  in a conventional
form
\[
U = U_0 - \frac{1}{2} m^2 \varphi^2 + \frac{\lambda}{4} \varphi^4.
\]
which implies  the so called `rolling down' of the field $\varphi$ from
its unstable value $\varphi=0$ with $U=U_0$  to stable asymptotic
value $\varphi_0$ which is defined from the equation $U(\varphi_0)=0$.
This potential, however, must explicitly satisfy the `initial' condition
$U \str 0$ as $a \str 0$. Assuming that the rolling down of the field
$\varphi$ starts at $a=0$ we find ourselves in a difficult situation
because the potential $U$ as a function of $\varphi$ must have a nonzero
value for the initial value $\varphi=0$ whereas
the same potential must be equal zero at $a=0$ as a function of $a$.

One possible solution of this problem is to take into account
that  the potential $U$ after solving the equation (\ref{potu})
is a function of $a$. Since we know {\it a priori} what properties
of the model to expect  we can postulate $U$ in the
form
\be
U = a^2 H^2 \lambda (\varphi^2  -  \varphi_0^2)^2
\label{pot**}
\ee
where $\lambda$ and $\varphi_0$ are constants. The theory with this
potential will also benefit from the fact that the
dynamical equation (\ref{potu}) can be solved in terms of the
variable  $a$ without knowledge of $H(a)$.

This potential satisfies naturally the condition $U \str 0$ as $a \str 0$.
From (\ref{pot**}) one can intuitively conclude that the field
$\varphi$ evolves
to $\pm \varphi_0$ which is the turning point of the potential.
The dimensionless  constant  $\varphi_0$ is a free parameter
and corresponds to the value of the field at
stable vacuum with zero energy. It is clear that the potential
(\ref{pot**}) has a maximum at $\varphi = 0$, so that one can argue
that the value of $U_0 = \lambda\varphi_0^4$  determines a temporal
scale (in terms of $a$) of decay  of the unstable vacuum  into
particles as it is in a standard inflationary scenario. We show below
that the parameter $\left( 2 \sqrt{V_0} \right)^{-1} $ can be interpreted
geometrically as the value of the scale factor at the point of
exit from inflation.

The equation (\ref{potu}) now has the form
\[
\pdp = 4 \lambda (\varphi^2  -  \varphi_0^2)^2
\]
and its solution is
\[
\varphi = \varphi_0 \left[ \frac{1 + C \exp{\pm 4 \srl \varphi_0 a}}
                                {1 - C \exp{\pm 4 \srl \varphi_0 a}}
                     \right].
\]
From the asymptotic condition $\varphi \str \varphi_0$ as $a \str \infty$
we choose the sign ($-$) in the exponential functions; then from the
initial condition  $\varphi(a=0) = 0$ one finds $C=-1$ so that finally
\be
\varphi = \varphi_0 \left[ \frac{ 1 -  \exp (- 4 \srl \varphi_0 a) }
                                { 1 +  \exp (- 4 \srl \varphi_0 a) }
                     \right] \, = \, \varphi_0 \, \tanh [ 2 \, \srl \,
                     \varphi_0 \, a]
\label{solphi}
\ee

This solution describes a standard rolling down of the field $\varphi$
along the potential curve $U$ from the unstable value $\varphi =0$
at $a=0$  (where $\pdp \not = 0$) to the stable state with
$\varphi = \varphi_0$ and zero potential.
From (\ref{solphi}) one can find
\be
\pds =  4 H^2 \left[\frac{a}{a_{*}} \right]^2
\cosh^{-4} \left[\frac{2}{\varphi_0}\frac{a}{a_{*}} \right],
\label{pdsa}
\ee
where we introduced a parameter $a_{*} \equiv
( \varphi_0^2 \srl )^{-1}$.
The field  $\varphi$  evolves effectively to its  stable value
$\varphi_0$ during the interval
$(a_{*} \varphi_0)/2$. This implies that the contribution of
$\pds$ to $\rr$ \ (\ref{rad}) and $N$ \ (\ref{num}) is
exponentially small for $a > a_{*} \varphi_0$.
\section{The Hubble parameter $\mathbf H(a)$}                               %
Now we are in a position to make a prediction of the asymptotic
form of the Hubble parameter $H(a)$.

Making  a natural  physical assumption
that $N(a) \str N_{\infty} < \infty$
as $a \str \infty$ (in analogy with \cite{gmn}),
one can obtain from (\ref{num}) the asymptotic
for $H(a)$ as $a \gg a_{*} \varphi_0$ neglecting  the contribution
from $\pds$ in the formula for $N$ (\ref{num}):
\be
H(a) \sim  a^{-2} \qquad  {\rm as } \qquad a \str \infty.
\label{asinfty}
\ee
In the region $a \ll (a_{*} \varphi)/2 $ one can use an approximate
expression for $\pds$
\[
\pds \approx \,4  H^2 \left[\frac{a}{a_{*}} \right]^2,
\]
which, being substituted into the equation  (\ref{num}) leads
to the differential equation for $H^2(a)$:
\be
[H^2(a)]' +  \frac{4a}{a_{*}^2} H^2(a) = - C \frac{N^{4/3}(a)}{a^5}.
\label{eqH}
\ee
In the limiting case $N(a) \str 0$ as $a \str 0$ one can obtain a formal
nontrivial solution of this equation
\[
H^2(a) \sim
           D \,  \exp{(- 2 \left[\frac{a}{a_{*}} \right]^2)}.
\]
which gives an accurate  asymptotic for $H^2$ as $a \str 0$:
\be
H^2(a) \approx
           D\, \left( 1 - 2 \, \left[\frac{a}{a_{*}} \right]^2 \right)
\qquad {\rm as } \qquad a \str 0 \,.
\label{aszero}
\ee
Comparing (\ref{asinfty}) and (\ref{aszero}) one can argue
that the simplest smooth
function $H(a)$ satisfying both these asymptotic conditions is
\[
H(a) = \frac{C}{a_*^2 + a^{2}},
\]
where $C, B$ are constants. This describes a universe
with a smooth transition from inflation to radiation. The universe
is initially de Sitter-like (since $H \approx $ constant
for small $a$), and becomes radiationlike (since $H \sim 1/a^2$
for large $a$).
The parameter $a_*$ can have a geometrical interpretation now as
a value of the scale factor
at exit from inflation, defined in general geometrically from
the condition  $\ddot{a}(t_{\rm e}) = 0$, or equivalently
$H_e = - a_e H_e'$,  i.e.
\[
a_* = a_{\rm e} \equiv a(t_{\rm e}).
\]
Thus the form of the Hubble rate is
\begin{equation}
H(a) =
2H_{\rm e}\left({a_{\rm e}^2\over a_{\rm e}^2 + a^2}\right)\,.
\label{hubble}
\end{equation}
This form
was  presented in \cite{roydavid} as an ad hoc prescription
to achieve smooth exit from inflation to radiation, but without
a physical basis such as that given here. In the paper
\cite{gmn} we obtained the same form of
the Hubble rate in the model with a smooth exit from inflation
as an exact solution of a differential equation subject to
a hypothesis on the rate of particle creation
in the model. The novelty of this paper is that it is possible  to
link the parameter of exit from inflation $a_e$ with the parameters
of the self-interacting potential of the scalar field $\varphi$:
\be
a_{\rm e} = a_*  =   ( \varphi_0^2 \srl )^{-1}.
\label{exit}
\ee

The expression for the
cosmic proper time follows on integrating Eq. (\ref{hubble})
(see also \cite{gmn}):
\[
t=t_{\rm e}+{1\over 4H_{\rm e}}\left[
\ln\left({a\over a_{\rm e}}\right)^2+
\left({a\over a_{\rm e}}\right)^2-1\right]\,.
\]
\section{Thermodynamics of radiation}
On substituting now the expression for $\pds$ (\ref{pdsa})
with $a_* = a_e$ into the formula for the energy density of radiation
(\ref{rad}), one obtains
\be
\rr(a) = \rho_r^0(a) \left[1 - F\left(\frac{a}{a_{\rm e}}, \varphi_0\right) \right],
\label{enrad}
\ee
where
\be
\rho_r^0(a) =  12 H_{\rm e}^2 \left(\frac{a}{a_{\rm e}} \right)^2
{ \left( \frac{a_{\rm e}^2}{a_{\rm e}^2 + a^2} \right)^3 } \,
\ee
is the energy density of radiation in the case of $\varphi \equiv 0$
(compare with \cite{gmn}), and
\be
F\left(\frac{a}{a_{\rm e}}, \varphi_0\right) =
\left[1+ \left(\frac{a}{a_{\rm e}} \right)^2 \right]
\cosh^{-4} \left(\frac{2}{\varphi_0}\frac{a}{a_{\rm e}} \right).
\ee
From the physical condition  $\rr \geq 0$,  and hence from $F \leq 1$,
one can find the range of values for $\varphi_0$:
\be
0 \leq \varphi_0 \leq 2 \sqrt{2},
\label{range}
\ee
which in conjunction with (\ref{exit}) determines the range of possible
values of $\lambda$ (subject to our knowledge of $a_{\rm e}$).
It is clear that  $F(a=0)=1$. It follows from (\ref{solphi})
that  for all possible values of $\varphi_0$ from (\ref{range})
the field $\varphi$ decays on scales $a<a_e$, so that
it is impossible to extend the decay of $\varphi$ for  $a > a_e$.

For the energy density of the field $\varphi$ we have
\be
\rp (a) =  \Lambda (a) \left[ 1  +  \left(\frac{a}{a_{\rm e}} \right)^2
F\left(\frac{a}{a_{\rm e}}, \varphi_0\right)   \right],
\label{enfield}
\ee
where
\be
\Lambda (a) = 12H_{\rm e}^2\left({a_{\rm e}^2\over
a_{\rm e}^2+a^2}\right)^3 \,.
\label{rholam}
\end{equation}
is the $\rp$ when $\varphi \equiv 0$ i.e. the decaying cosmological
constant of the paper \cite{gmn}.
It follows that $\rho_{\varphi}(0)  = 12 H_{\rm e}^{2}$. Note that
(\ref{hubble}) implies $H(0)=2H_{\rm e}$.

The function $F$ represents a rapid exponential decay so that
one can effectively treat it as zero for $a>a_{\rm e}$. This gives
the expected asymptotics for $\rr$  and $\rp$ as $a \str \infty$:
\[
\rho (a) \sim \frac{1}{a^4}\,,~~~~ \rp(a) \sim \frac{1}{a^6}\,,
\]
so that $\rp$ rapidly becomes negligible in comparison with $\rr$.
\vspace{1ex}

Comparing the functions $\rho_r^0, \, \rr,
\Lambda, \, \rp$ in terms of the dimensionless variable
$x=a/a_{\rm e}$  one can conclude that
the contribution of $\pds$ in the expressions for $\rr$ and
$\rp$, (which appears in the formulas (\ref{enrad}) and (\ref{enfield})
through the function $F$), is noticeable only for $a<a_{\rm e}$. The order
of magnitude for $\rr$ and $\rho_r^0$, and for $\rp$ and $\Lambda$
correspondingly is the same, leading only to some shift of the point
of maximum for $\rp$ towards exit from inflation. Note, that
$\rho_r^0$ reaches a maximum at $a_{\rm m} = a_{\rm e}/\sqrt{2}$, with
\[
\rho_{\rm m}^0 \equiv \rho^0(a_{\rm m}) =
{\textstyle{16\over9}}\, H_{\rm e}^{2}\,,
~~~~~ \Lambda (a_{\rm m}) = 2 \rho_{\rm m}^0 \,.
\]
Note also from the graph that $\rr$ and $\Lambda$ are equal at exit
up to terms of $O(10^{-2})$:
\[
\rr(a_{\rm e}) \approx \rr^0(a_{\rm e}) =
{\textstyle{3\over2}} H_{\rm e}^{2} = \Lambda (a_{\rm e})
\approx \rp(a_{\rm e})
 \,,
\]
while $\rp \ll \rr$ for $a \gg a_{\rm e}$, i.e. during
the radiation-dominated era.

The formulas (\ref{enrad}) and (\ref{enfield}) reflect the
creation of radiation due to vacuum decay.
The initial value $\rr(0) = 0$ confirms that the field
corresponding to radiation is initially in a regular vacuum state.

Substituting the equation (\ref{hubble})
into equation (\ref{num}) we get the exact form for the
particle number
\begin{equation}
N(a) = N^0(a)\,\left[1 - F\left(\frac{a}{a_{\rm e}}, \varphi_0\right)
\right]^{3/4}
\label{number}
\end{equation}
where
\[
N^0(a) = N_{\infty} \left(\frac{a^2}{a_{\rm e}^2 + a^2}\right)^{9/4}\,,
\]
is the particle number in the case $\varphi=0$, and  $N_{\infty}$
is usually taken to be about $10^{88}$ (see \cite{gmn}).
It is clear that the
presence of $F$ in the formula (\ref{number}) does not
affect the asymptotic behavior of $N$ as $a \str \infty$, dying
away exponentially on scales of the order of $a_{\rm e}$.
It means that after exit from inflation the behavior of radiation
and the decay of the  energy density for $\varphi$ will be
practically  the same as they were in the case of the cosmological
constant decay  \cite{gmn}.

Since $N(0)= n(0) = 0$,
the initial state of the field has no particles, i.e. it is a
regular vacuum. The number density is
\begin{equation}
n(a) = \cosh 2 \, 2^{5/4} \, n_{\rm e}\left({a_{\rm e}\over a}\right)^3
\left(\frac{a^2}{a_{\rm e}^2
+ a^2}\right)^{9/4} (1-F)^{3/4}\,.
\label{n}\end{equation}


In analogy with \cite{gmn}  it seems reasonable to use the
black-body relation
for the radiation throughout the expansion, and to define the
temperature by
\begin{equation}
T(a) = \frac{1}{3}\frac{\rho(a)}{n(a)} =
{H_{\rm e}^2  2^{3/4}\over n_{\rm e} \cosh 2}
\left({a_{\rm e}\over a}\right)\left({a^2
\over a_{\rm e}^2+a^2}\right)^{3/4} (1-F)^{1/4} \,,
\label{temperature}
\end{equation}
where we have used equations
(\ref{enrad}) and (\ref{n}).
At the initial radiation vacuum, it is clear that
$T(0)=0$.
During the radiation era, i.e. for $a \gg a_{\rm e}$,
\[
T \sim a^{-1}\,,
\]
in agreement with the standard result for free radiation in
an expanding universe.

The formulas for $\rho$ and $n$ can be presented in the thermodynamic
form
\begin{equation}
\rho = 24\left(\frac{ n_{\rm e}^4}
{H_{\rm e}^6}\right) T^{4}\,,~~~~~
n = 8\left(\frac{ n_{\rm e}^4}{H_{\rm e}^6}\right) T^{3}\,.
\label{tdform}
\end{equation}
Combining now the Gibbs equation
\[
T dS = d(\rho V) + p d V\,,
\]
with equation (\ref{firstlaw}), and using the definition
(\ref{temperature}) of $T$, we obtain the
entropy of radiation in the observable universe as
\begin{equation}
S (a) = 4 N (a)\,,
\label{entropy}
\end{equation}
and leads to a value
of the same order of magnitude as our result.
\section{ Conclusion}  %
One can conclude that the presence of the scalar field $\varphi$
in this model does not change considerably the physical results which
have been obtained in the paper  \cite{gmn} with the
decaying cosmological constant.
The small difference with the paper \cite{gmn}
can be observed only for $a<a_e$.
One can not extend the decay of the field $\varphi$ beyond
$a=a_{\rm e}$ in this model because of the condition on the range of values
of $\varphi_0$ \ (\ref{range}). This means that the
thermodynamic analysis for $a>a_e$ is similar to the paper
 \cite{gmn} and gives the same prediction for the
total entropy produced in the universe.

Generalizing the results of this paper one can claim that
models with  decaying cosmological constant $\Lambda$
corresponding to a special case of the equation of state
$\pp = - \rp$,   describe adequately the smooth transition
from inflation to radiation and give a reasonable prediction
for the entropy of matter in the universe.



\begin{thebibliography}{99}
\bibitem{gmn}   Gunzig E., Maartens R., Nesteruk  A. V. (1998)
                {\em Class. Quant. Grav.} {\bf 15} 923.


\bibitem{hupark77}  Hu, B. L. and Parker, L. (1977) {\em Phys. Lett.} A
                {\bf 63} 217.

\bibitem{creation} Hu, B. L. (1983) {\em Phys. Lett.} A {\bf 97} 368;
                Birrell N D and Davies P. C.  W. (1982)
                {\it Quantum Fields in Curved Space}
                (Cambridge: Cambridge University Press);
                Allen, B. (1988) {\em Phys. Rev.} D {\bf 37} 2078;
               Linde, A. D. 1990 {\it Particle Physics and
                Inflationary Cosmology} (Geneva: Harwood Academic);
                Nesteruk, A. V. and Ottewill, A. C. (1995) {\em Class.
                Quantum Grav.} {\bf 12} 51.
                See also Lyth, D. H. and Roberts, D. 1996 {\em Preprint}
                hep-ph/9609441,  and the papers cited there.

\bibitem{edgard1} Prigogine, I., Geheniau, J., Gunzig, E. and
                  Nardone, P. (1989) {\em Gen. Rel. Grav.} {\bf 21}
                  767;
                  Gunzig, E. and Nardone, P. (1989) {\em Int. J. Theor.
                  Phys.} {\bf 28} 927.

\bibitem{lima1}  Abramo, L. R. W. and Lima, J. A. S. (1996) {\em Class. Quantum
                 Grav.} {\bf 13} 2953

\bibitem{lima2}  Lima, J. A. S. (1996) {\em Phys. Rev.} D {\bf 54} 2572.

\bibitem{vl}
Viana, P. T. P. and Liddle, A. R. (1997) {\em Phys. Rev.} D, in press
(astro-ph/9708247);
Coble K., Dodelson, S. and Frieman, J. A. (1997) {\em Phys. Rev.} D
{\bf 55} 1851.

\bibitem{mubarozer} Mubarak K. M., \"Ozer M. (1998)
                {\em Class. Quant. Grav.} {\bf 15} 75.

\bibitem{ozertaha} \"Ozer M. , Taha M. O. (1998)
           {\em Los Alamos Preprint astro-ph/9802023}.

\bibitem{overcoop} Overduin J.M., Cooperstock F.I. (1998)
           {\em Los Alamos Preprint astro-ph/9805260}.


\bibitem{krausturner}  L. M. Krauss and M. S. Turner, (1995)
                {\em Gen. Rel. Grav.} {\bf 27}, 1137;

\bibitem{qft} P.G.Bergmann, (1968)
                Int. J. Theor. Phys. {\bf 1}, 25;
                R.V.Wagoner, (1970)
                Phys. Rev. {\bf D1}, 3209;
                A.D.Linde, (1974)
                JETP Letters {\bf 19}, 183;
                M.End\={o} and T.Fukui, (1977)
                Gen. Rel. Grav. {\bf 8}, 833;
                V.Canuto, S.H.Hsieh and P.J.Adams, (1977)
                Phys. Rev. Lett. {\bf 39}, 429;
                D.Kazanas, (1980)
                Astrophys. J. Lett. {\bf 241}, L59.
                A.M.Polyakov, (1982)
                Sov. Phys. Usp. {\bf 25}, 187;
                S.~L.~Adler, (1982)
                Rev. Mod. Phys. {\bf 54}, 729.

\bibitem{Scalar} A. D. Dolgov, in
                The Very Early Universe, ed.
                G. W. Gibbons, S. W. Hawking and S.T.C.Siklos
                (Cambridge: Cambridge University Press, 1983), p. 449;
                 L. F. Abbott, (1985)
                 Phys. Lett. {\bf 150B}, 427;
                 T. Banks, (1985)
                 Nucl. Phys. {\bf B249}, 332;
                 R. D. Peccei, J. Sol\`a and C. Wetterich, (1987)
                 Phys. Lett. {\bf 195B}, 183 ;
                 S. M. Barr, (1987)
                 Phys. Rev. {\bf D36}, 1691;
                 P.J.E. Peebles and B. Ratra, (1988)
                 Astrophys. J. Lett. {\bf 325}, L17;
                 Y. Fujii and T. Nishioka,  (1991)
                Phys. Lett. {\bf 254B}, 347 ;
                 J. A. Frieman {\em et al\/}, (1995)
                 Phys. Rev. Lett. {\bf 75}, 2077;
                 J. W. Moffat, (1995)
                 Phys. Lett. {\bf B357}, 526.
\bibitem{wei89} S. Weinberg, (1989)
                Rev. Mod. Phys. {\bf 61}, 1.

\bibitem{parker} Parker l.  (1998)
           {\em Los Alamos Preprint gr-qc/9804002}.


\bibitem{kolb&turner90} Kolb, E. W. and Turner, M. S.  (1990) {\it The Early
                             Universe} (New York: Addison-Wesley);
          Liddle, A. R. and Lyth, D. H. (1993) {\em Phys. Rep.} {\bf 231} 1

\bibitem{yoko}Yokoyama, J., Maeda, K. (1988) {\em Phys. Lett.}  {\bf
B 207} 31

\bibitem{oliveira} De Oliveira, H. P., Ramos, R. O. (1997)
           {\em Los Alamos Preprint gr-qc/9710093}.

\bibitem{roydavid} Maartens,  R., Taylor, D. R. and Roussos, N. (1995)
               {\em Phys. Rev.} D {\bf 52} 3358

\bibitem{alexei_96} Nesteruk, A. V. (1996) {\em Europhys. Lett.}
                                       {\bf 36} 233

\bibitem{caldwell} Caldwell, R. R. (1996) {\em Class. Quantum Grav.}
{\bf 13} 2437

\bibitem{b1} Berera, A. (1997) {\em Phys. Rev.} D {\bf 55} 3346

\end{thebibliography}
\end{document}